\documentclass[12]{article}
\usepackage{fleqn}
\usepackage{graphicx}
\usepackage{amssymb}
\begin{document}
\Large
\begin{center}
{\bf Geometric Hyperplanes of the Near Hexagon\\ L$_3$\,$\times$\,GQ(2,\,2)}
\end{center}
\large
\vspace*{.2cm}
\begin{center}
M. Saniga$^{1,5}$,  P. L\' evay$^{2,5}$, M. Planat$^{3,5}$ and P. Pracna$^{4,5}$
\end{center}
\vspace*{-.1cm} \normalsize
\begin{center}
$^{1}$Astronomical Institute, Slovak Academy of Sciences\\
SK-05960 Tatransk\' a Lomnica, Slovak Republic\\
(msaniga@astro.sk)

\vspace{0.3cm}

$^{2}$Department of Theoretical Physics, Institute of Physics\\
Budapest University of Technology and Economics, H-1521 Budapest, Hungary\\
(levay@neumann.phy.bme.hu)

\vspace{0.3cm}

$^{3}$Institut FEMTO-ST, CNRS, 
32 Avenue de
l'Observatoire\\ F-25044 Besan\c con Cedex, France\\
(michel.planat@femto-st.fr)

\vspace{0.3cm}

$^{4}$J. Heyrovsk\' y Institute of Physical Chemistry, v.v.i., Academy of Sciences of the Czech Republic,
Dolej\v skova 3, CZ-182 23 Prague 8, Czech Republic\\
(pracna@jh-inst.cas.cz)

\vspace*{.3cm}

$^{5}$Center for Interdisciplinary Research (ZiF), University of Bielefeld\\
D-33615 Bielefeld, Germany

\vspace*{.3cm}

(1 September 2009)

\end{center}

\vspace*{-.3cm} \noindent \hrulefill

\vspace*{.0cm} \noindent {\bf Abstract}\\
Having in mind their potential quantum physical applications, we classify all geometric hyperplanes 
of the near hexagon that is a direct product of a line of size three and the generalized quadrangle 
of order two. There are eight different kinds of them, totalling to $1023 = 2^{10} - 1$ = $|$PG$(9, 2)|$, 
and they form two distinct families intricately related with the points and lines of the Veldkamp space of the 
quadrangle in question.

\noindent 
\\ 
{\bf MSC Codes:} 51A99, 51E12, 51E30\\
{\bf Keywords:}  Near Hexagons -- Geometric Hyperplanes -- Veldkamp Spaces -- Qubits

\vspace*{-.2cm} \noindent \hrulefill

\vspace*{.1cm}
\section{Introduction}
There are quite a few finite geometries/point-line incidence structures that have recently been recognized to play an important role in physics.
Amongst them, the one that acquired a particular footing is GQ$(2, 2)$ --- the unique generalized quadrangle of order two. On its own, the GQ$(2, 2)$
is the underlying framework for fully expressing  the commutation relations between the elements of a two-qubit generalized Pauli group in geometrical terms \cite{spp,ps2}.
As a subgeometry/subconfiguration, the GQ$(2, 2)$ underpins a particular kind of truncation of the $E_{6(6)}$-symmetric black hole/black string entropy formula in five dimensions \cite{lsvp}.
In both the cases, remarkably, it is also the geometric hyperplanes of the GQ$(2, 2)$ that enter the game in an essential way.

In view of these developments, it is likely that there are other physically relevant finite geometries incorporating GQ$(2, 2)$s. In this paper we shall have a look at the most
promising candidate of them: the slim dense near hexagon that originates as a direct product of a projective line over the field of two elements and a  GQ$(2, 2)$, together with the totality of its geometric hyperplanes.

\section{Generalized Quadrangles, Near Polygons, Geometric Hyperplanes and Veldkamp Spaces}
We start with a brief overview of the essential theory and nomenclature; for more details, the interested reader is referred to \cite{pay-thas}--\cite{buek}.

A {\it finite generalized quadrangle} of order $(s, t)$, usually denoted GQ($s, t$), is an incidence structure $S = (P, B, {\rm I})$,
where $P$ and $B$ are disjoint (non-empty) sets of objects, called respectively points and lines, and where I is a symmetric point-line
incidence relation satisfying the following axioms \cite{pay-thas}: (i) each point is incident with $1 + t$ lines ($t \geq 1$) and two
distinct points are incident with at most one line; (ii) each line is incident with $1 + s$ points ($s \geq 1$) and two distinct lines
are incident with at most one point;  and (iii) if $x$ is a point and $L$ is a line not incident with $x$, then there exists a unique
pair $(y, M) \in  P \times B$ for which $x {\rm I} M {\rm I} y {\rm I} L$; from these axioms it readily follows that $|P| = (s+1)(st+1)$
and $|B| = (t+1)(st+1)$. It is obvious that there exists a point-line duality with respect to which each of the axioms is self-dual.
If $s = t$,
$S$ is said to have order $s$. The generalized quadrangle of order $(s, 1)$ is called a grid and that of order $(1, t)$ a dual grid. A
generalized quadrangle with both $s > 1$ and $t > 1$ is called thick.
Given two points $x$ and $y$ of $S$ one writes $x \sim y$ and says that $x$ and $y$ are collinear if there exists a line $L$ of $S$
incident with both. For any $x \in P$ denote $x^{\perp} = \{y \in P | y \sim x \}$ and note that $x \in x^{\perp}$;  obviously, $x^{\perp}
= 1+s+st$.  A triple of pairwise non-collinear points of $S$ is called a {\it triad}; given
any triad $T$, a point of $T^{\perp}$ is called its center and we say that $T$ is acentric, centric or unicentric according as
$|T^{\perp}|$ is, respectively, zero, non-zero or one. An ovoid of a generalized quadrangle $S$ is a set of points of $S$ such that each
line of $S$ is incident with exactly one point of the set;  hence, each ovoid contains $st + 1$ points.

A {\it near polygon} (see, e.\,g., \cite{bruyn} and references
therein) is a connected partial linear space $S = (P, B, {\rm I})$, I $
\subset P \times L$, with the property that given a point $x$ and
a line $L$, there always exists a unique point on $L$ nearest to
$x$. (Here distances are measured in the point graph, or
collinearity graph of the geometry.)  If the maximal distance
between two points of $S$ is equal to $d$, then the near polygon
is called a near $2d$-gon. A near 0-gon is a point and a near
2-gon is a line; the class of near quadrangles coincides with the
class of generalized quadrangles.
A nonempty set $X$ of points in a near polygon $S = (P, B, {\rm I})$ is
called a subspace if every line meeting $X$ in at least two points
is completely contained in $X$. A subspace $X$ is called
geodetically closed if every point on a shortest path between two
points of $X$ is contained in $X$. Given a subspace $X$, one can
define a sub-geometry $S_X$ of $S$ by considering only those
points and lines of $S$ which are completely contained in $X$. If
$X$ is geodetically closed, then $S_X$ clearly is a
sub-near-polygon of $S$. If a geodetically closed sub-near-polygon
$S_X$ is a non-degenerate generalized quadrangle, then $X$ (and
often also $S_X$) is called a {\it quad}.

A near polygon is said to have order $(s, t)$ if every line is
incident with precisely $s+1$ points and if every point is on
precisely $t+1$ lines. If $s = t$, then the near polygon is said
to have order $s$. A near polygon is called {\it dense} if every
line is incident with at least three points and if every two
points at distance two have at least two common neighbours. A near
polygon is called {\it slim} if every line is incident with
precisely three points. It is well known (see, e.\,g.,
\cite{pay-thas}) that there are, up to isomorphism, three slim
non-degenerate generalized quadrangles. The $(3 \times 3)$-grid is
the unique generalized quadrangle GQ$(2, 1)$.
The unique generalized quadrangle GQ$(2, 2)$, often dubbed the doily, is the
generalized quadrangle of the points and those lines of PG(3,\,2) which
are totally isotropic with respect to a given symplectic polarity.
The points and lines lying on a given nonsingular elliptic quadric
of PG$(5, 2)$ define the unique generalized quadrangle GQ$(2, 4)$. Any {\it slim dense} near polygon contains
quads, which are necessarily isomorphic to either GQ$(2, 1)$,
GQ$(2, 2)$ or GQ$(2, 4)$.

The incidence structure $S$ is called the direct product of $S_1$
and $S_2$, and denoted by $S_1 \times S_2 (\simeq S_2 \times S_1)$, if: i) $P := P_1 \times P_2$;
ii) $B := (P_1 \times B_2) \cup (B_1 \times P_2)$; and iii) the point $(x, y)$ of
$S$ is incident with the line $(z, L) \in P_1 \times B_2$ if and only if $x=z$ and
$y {\rm I}_2 L$ and with the line $(M, w) \in B_1 \times P_2$ if and only if $x {\rm I}_1 M$ and  $y=w$.

Next, a {\it geometric hyperplane} of a partial linear space is a
proper subspace meeting each line (necessarily in a unique point
or the whole line). 
For $S =$ GQ($s,
t$), it is well known that $H$ is one of the following three
kinds: (i) the perp-set of a point $x$,  $x^{\perp}$; (ii) a
(full) subquadrangle of order ($s,t'$), $t' < t$; and (iii) an
ovoid.
The set of points at non-maximal distance from
a given point $x$ of a dense near polygon $S$ is a hyperplane of
$S$, usually called the {\it singular} hyperplane with {\it
deepest} point $x$. Given a hyperplane $H$ of $S$, one defines the
{\it order} of any of its points as the number of lines through
the point which are fully contained in $H$; a point of a
hyperplane/sub-configuration is called {\it deep} if all the lines
passing through it are fully contained in the
hyperplane/sub-configuration. If $H$ is a hyperplane of a dense
near polygon $S$ and if $Q$ is a quad of $S$, then precisely one
of the following possibilities occurs: (1) $Q \subseteq H$; (2) $Q
\cap H = x^{\perp} \cap Q$ for some point $x$ of $Q$; (3) $Q \cap
H$ is a sub-quadrangle of $Q$; and (4) $Q \cap H$ is an ovoid of
$Q$. If case (1), case (2), case (3), or case (4) occurs, then $Q$
is called, respectively, {\it deep}, {\it singular}, {\it
sub-quadrangular}, or {\it ovoidal} with respect to $H$. 

Finally, we shall introduce the notion of the {\it Veldkamp
space} of a point-line incidence geometry $S(P, B, {\rm I})$,
$\mathcal{V}(S)$ \cite{buek}, which is the
space in  which (i) a point is a geometric hyperplane of  $S$
and (ii) a line is the collection 
$H'H''$ of all geometric
hyperplanes $H$ of $S$  such that $H' \cap H'' = H' \cap H =
H'' \cap H$ or $H = H', H''$, where $H'$ and $H''$ are distinct
points of $\mathcal{V}(S)$.

\section{L$_3$\,$\times$\,GQ(2,\,2) and its Geometric Hyperplanes}
\subsection{L$_3$\,$\times$\,GQ(2,\,2)}
The unique point-line incidence geometry L$_3$\,$\times$\,GQ(2,\,2) is obtained by taking three isomorphic copies
of the generalized quadrangle GQ(2,\,2) and joining the corresponding points
to form lines of size 3. It is a slim dense near hexagon having 45 points and 60 lines, with four lines through a point.
The number of common neighbours of two points $x, y$ at distance two, that is $|x^{\perp} \cap y^{\perp} |$, is either two or three. L$_3$\,$\times$\,GQ(2,\,2) contains 15 GQ$(2, 1)$-quads (henceforth simply grid-quads) and
three GQ$(2, 2)$-quads (doily-quads). The lines of L$_3$\,$\times$\,GQ(2,\,2) are of two distinct types according as they lie in three grid-quads (type one) or a grid-quad and a doily-quad (type two); 
there are 15 lines of type one and 45 of type two, with each point being on one line of type one and three lines of type two. 
Also, each grid-quad features both kinds of lines in equal proportion, whereas a doily-quad consists solely of type-two lines.
This near hexagon
can be universally embedded in PG$(9, 2)$ and its full group of automorphisms is isomorphic to $S_6 \times S_3$ \cite{bchw,ron}.
The structure of L$_3$\,$\times$\,GQ(2,\,2) is fully encoded in the properties of its geometric hyperplanes, which we will now focus on.

\begin{figure}[t]
\centerline{\includegraphics[width=\textwidth,clip=]{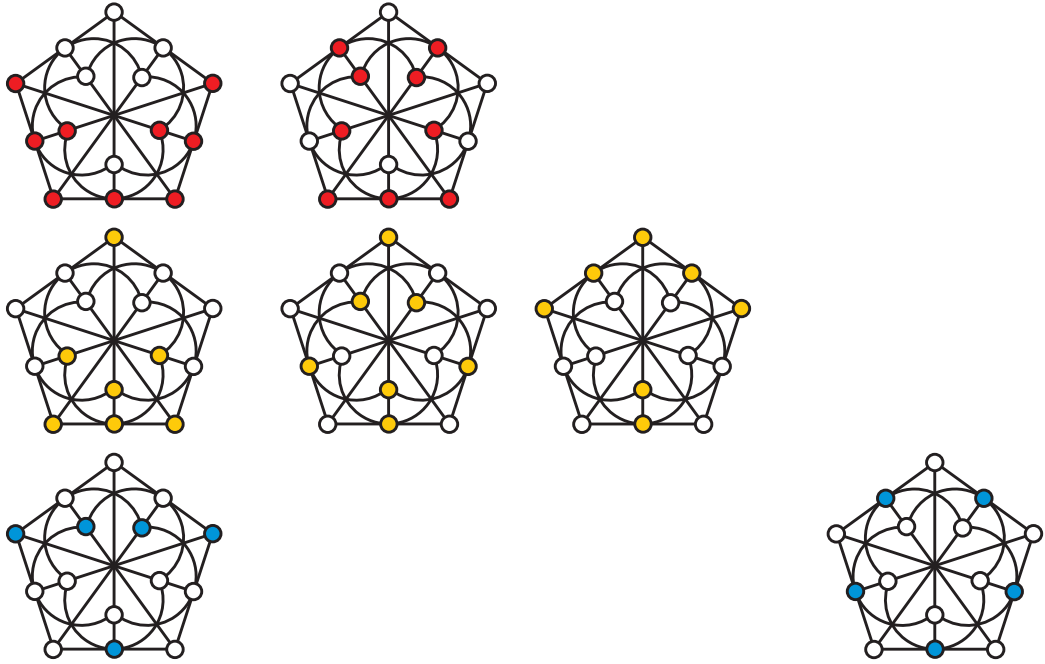}}
\vspace*{.2cm}
\caption{The three kinds of geometric hyperplanes of GQ(2,\,2). The points of the quadrangle are represented by small circles and its lines are illustrated by the straight segments as
well as by the segments of circles; note that not every intersection of two segments counts for a point of
the quadrangle. The upper panel shows perp-sets (yellow bullets), the middle panel grids (red bullets) and the bottom panel ovoids (blue bullets). Each picture --- except that in the bottom right-hand corner ---
stands for five different hyperplanes, the four other being obtained from it by its successive rotations through 72 degrees around the center of the pentagon.}
\end{figure}

\subsection{The Veldkamp Space of GQ(2,\,2)}
To this end, we shall first recall basic properties of the Veldkamp space of the doily, $\mathcal{V}$(GQ(2,\,2)) $\simeq$ PG(4,\,2), whose in-depth description can be found in \cite{spph}. 
The 31 points of $\mathcal{V}$(GQ(2,\,2)), that is the 31 distinct copies of geometric hyperplanes of  GQ(2,\,2), are of three distinct types: 15 perp-sets, 10 grids and five ovoids --- as illustrated in Figure 1.
The 155 lines of $\mathcal{V}$(GQ(2,\,2)), each being of the form $\{H', H'', \overline{H' \Delta H''}\}$ where $H'$ and $H''$ are two distinct geometric hyperplanes and $\overline{H' \Delta H''}$
is the complement of their symmetric difference, split into five distinct types as summarized in Table 1 and depicted in Figure 2.

\begin{table}[pth!]
\begin{center}
\caption{A succinct summary of the properties of the five different types of the lines of $\mathcal{V}$(GQ(2,\,2))
in terms of the core (i.\,e., the set of points common to all the three hyperplanes forming a line) and the types of geometric hyperplanes featured by a generic line of a given type.
The last column gives the total number of lines per each type.} \vspace*{0.4cm}
\begin{tabular}{||c|c|ccc|c||}
\hline \hline
Type & Core & Perps & Ovoids & Grids & $\#$ \\
\hline
I & Pentad & 1 & 0 & 2 & 45\\
II & Collinear Triple & 3 & 0 & 0 & 15 \\
III &Tricentric Triad & 3 & 0 &  0 & 20 \\
IV & Unicentric Triad & 1 & 1 &  1 & 60 \\
V & Single Point & 1 &  2 & 0 & 15 \\
\hline \hline
\end{tabular}
\end{center}
\end{table}

\begin{figure}[pth!]
\centerline{\includegraphics[width=9truecm,clip=]{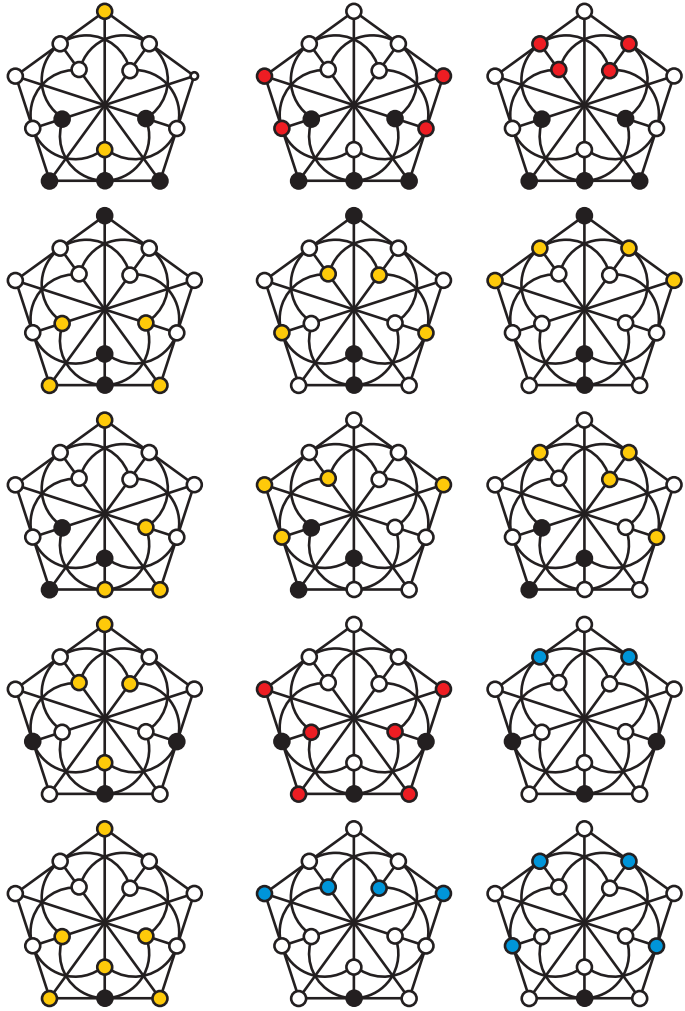}}
\vspace*{.2cm} \caption{The five different kinds of the lines of
$\mathcal{V}$(GQ(2,\,2)), each being uniquely determined by the
properties of its core (black bullets).}
\end{figure}

\begin{table}[h]
\begin{center}
\caption{An overview of the types of geometric hyperplanes of the near hexagon L$_3$\,$\times$\,GQ(2,\,2). For each type (Tp) of a hyperplane we give the number of points (Pt) and lines (Ln), followed by the cardinalities of the points
of a given order, cardinalities of deep (dp), singular (sg), ovoidal (ov) and subquadrangular (sq) quads of both kinds, and, finally, the total number of its copies (Cd).} \vspace*{0.3cm}
{\begin{tabular}{||c|c|c|c|c|c|c|c|c|c|c|c|c|c|c|c|r||} \hline \hline
\multicolumn{1}{||c|}{} & \multicolumn{1}{|c|}{} & \multicolumn{1}{|c|}{}  &  \multicolumn{5}{|c|}{}                        & \multicolumn{4}{|c|}{}             
& \multicolumn{4}{|c|}{}                             &\multicolumn{1}{|c||}{}\\
\multicolumn{1}{||c|}{} & \multicolumn{1}{|c|}{} & \multicolumn{1}{|c|}{}  &  \multicolumn{5}{|c|}{$\#$ of Points of Order} & \multicolumn{4}{|c|}{$\#$ of Grid-Quads}
&\multicolumn{4}{|c|}{$\#$ of Doily-Quads} & \multicolumn{1}{|c||}{}\\
 \cline{4-16}
Tp & Pt & Ln  & 0 & 1 & 2 & 3 & 4 & dp & sg & ov & sq & dp & sg & ov & sq &  Cd  \\
\hline \hline
$H_1$ & 33 & 36  & 0  & 0  &  0 &  24 &  9 & 6  &  9 &  0 & -- &  1 & 0 & 0 & 2 &   30 \\
$H_2$ & 29 & 28  & 0  & 0  & 12 &  8  &  9 & 3  & 12 &  0 & -- &  1 & 2 & 0 & 0 &   45 \\
$H_3$ & 25 & 20  & 0  & 10 &  0 &  10 &  5 & 0  & 15 &  0 & -- &  1 & 0 & 2 & 0 &   18 \\
\hline
$H_4$ & 25 & 20  & 0  & 2  & 12 &  10 &  1 & 2  &  9 &  4 & -- &  0 & 1 & 0 & 2 &   270 \\
$H_5$ & 21 & 12  & 0  & 12 & 6  &  0  &  3 & 1  &  6 &  8 & -- &  0 & 3 & 0 & 0 &   90 \\
$H_6$ & 21 & 12  & 0  & 9  & 9  &  3  &  0 & 0  &  9 &  6 & -- &  0 & 3 & 0 & 0 &   120 \\
$H_7$ & 21 & 12  & 2  & 6  & 9  &  4  &  0 & 0  &  9 &  6 & -- &  0 & 1 & 1 & 1 &   360 \\
$H_8$ & 17 & 4   & 8  & 8  & 0  &  0  &  1 & 0  &  3 & 12 & -- &  0 & 1 & 2 & 0 &   90 \\
\hline \hline
\end{tabular}}
\end{center}
\end{table}

\begin{figure}[pth!]
\centerline{\includegraphics[width=3.5cm,angle=0,clip=]{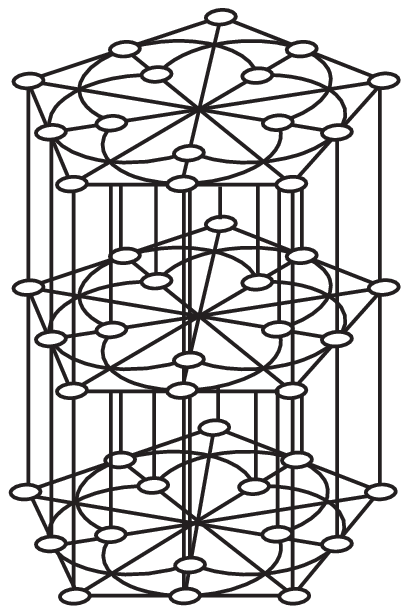}}
\vspace*{.2cm} \caption{A pictorial representation of L$_3$\,$\times$\,GQ(2,\,2). In this representation the three doily-quads lie in three parallel planes and the 15 lines of type one, 
``tying'' them together and lying fully in grid-quads, ``penetrate'' these planes perpendicularly.}
\end{figure}

\subsection{Geometric Hyperplanes of L$_3$\,$\times$\,GQ(2,\,2)}
Employing a ``cubic pentagon'' pictorial representation of L$_3$\,$\times$\,GQ(2,\,2) as shown in Figure 3, it was quite a straightforward task to find out all the types of geometric
hyperplanes of this geometry and to ascertain their basic characteristics, as summarized in Table 2. There are eight different kinds of them and they form two distinct families according as
they contain a deep doily-quad ($H_1$ to $H_3$) or not  ($H_4$ to $H_8$). The fine structure of the hyperplanes of the first family is given in Figure 4, that of the second family in Figure 5.
Comparing Figure 4 with Figure 1 and Figure 5 with Figure 2 one readily recognizes that this two-family split has a natural explanation in terms of the points and lines of $\mathcal{V}$(GQ(2,\,2)).

A hyperplane of the first family is always of such form that the two doily-quads which are not deep must not only contain the hyperplanes of the same type, but these must be joined by the same 
type-one lines. Since each of the three doily-quads can be deep, this implies $|H_1|$ =  3 $\times$ (the number of grids in GQ(2,\,2)) = 30, $|H_2|$ =  3 $\times$ (the number of perp-sets in GQ(2,\,2)) = 45 and $|H_3|$ =  3 $\times$ 
(the number of ovoids in GQ(2,\,2)) = 18; altogether, $|H_1| + |H_2| + |H_3|$ = 3 $\times$ (the number of Veldkamp {\it points} in the $\mathcal{V}$(GQ(2,\,2))) = 93.
On the other hand, there is an obvious one-to-one correspondence between the five types of hyperplanes of the second family and the five types of Veldkamp lines of $\mathcal{V}$(GQ(2,\,2)).

In the following, let $H$ be a hyperplane that contains no deep doily-quad. If the intersection of $H$ with two of the three doilies is given, the intersection of $H$ with the third is immediate. 
We use this to construct the remaining five types of geometric hyperplanes of L$_3\times$ GQ(2,\,2). It is straightforward to check supposing the contrary, that every hyperplane has a singular doily. So, let 
one doily be singular with deep point $P$. If a second doily is singular, too, with deep point $Q$, there arise two different kinds of hyperplanes according to whether the distance of 
$P$ and $Q$ is $2$ or $3$. If it is $2$, the third doily is singular, too, with deep point the third point of the unique triad in the grid-quad through $P$ and $Q$; this is type $H_5$. 
The number of type $H_5$ hyperplanes is $15 \times 6 = 90$. If it is $3$ and $P'$ and $Q'$ are the points of the third doily collinear with $P$, respectively $Q$, then no grid-quad is 
deep, the third doily is singular with deep point the third point of the triad through $P'$ and $Q'$; this is type $H_6$ and it holds $|H_6| = 15 \times 8 = 120$. Now suppose a second 
doily is subquadrangular. If the point $P'$ of the subquadrangular doily collinear with $P$ belongs to the subquadrangle, the third doily is subquadrangular, too; this is type $H_4$ which 
consists of $3 \times 15 \times 3 \times 2 = 270$ hyperplanes where $3$ possibilities for the singular quad with 15 different deep points exist, where three choices of subquadrangles 
through $P'$ exist and where two choices remain for subquadrangles in the third doily. If $P'$ does not belong to the subquadrangle, the third doily is ovoidal; this is type $H_7$ with 
$3\times 10 \times 2\times 6 = 360$ members where there are 3 choices for the subquadrangular doily with 10 possibilities for the choice of the grid, then 2 choices for the singular doily 
with 6 choices for the deep point of its singular hyperplane. If a second doily is ovoidal, we are back in type $H_7$ if the deep point of the singular quad is not collinear with any 
point of the ovoid of the ovoidal doily. If it is collinear with a point of the ovoid, then the third doily is ovoidal, too; this is type $H_8$ with 3 choices for the two ovoidal doilies, 
6 choices for the ovoid in one of the two ovoidal doilies and 5 possibilities for the deep point of the singular doily, altogether $3 \times 6 \times 5 = 90$. All in all, $|H_4| + |H_5| + \ldots 
+ |H_8|$ = 6 $\times$ (the number of Veldkamp {\it lines} in the $\mathcal{V}$(GQ(2,\,2))) = 930.

\begin{figure}[pth!]
\centerline{\includegraphics[width=\textwidth,clip=]{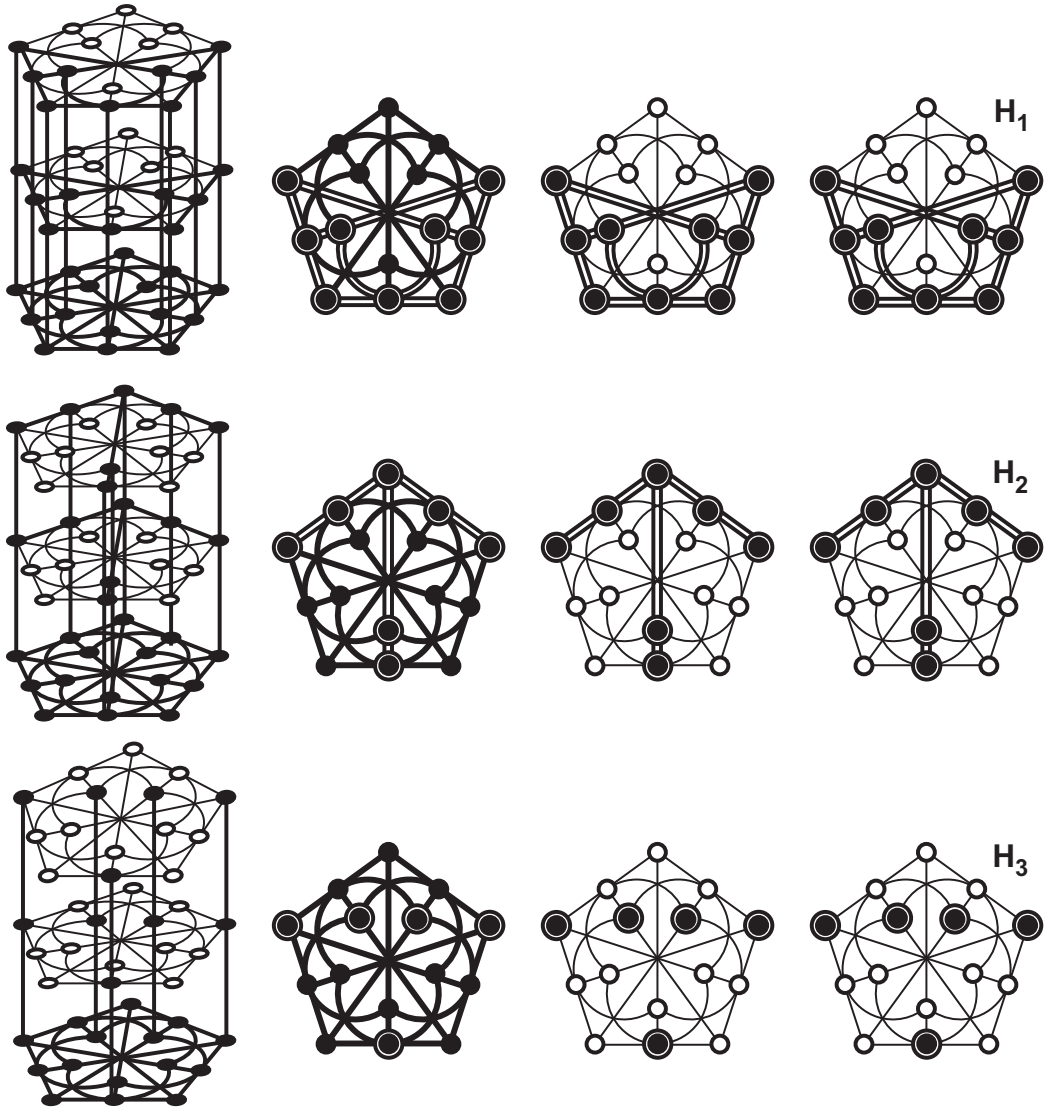}}
\vspace*{.8cm}
\caption{A diagrammatic illustration of the composition of the three kinds of geometric hyperplanes of the first family (see Table 2). In each row, the first picture of the four shows a compact, cubic pentagonal 
view of the hyperplane; to avoid its too crowded appearance, only those type-one lines are drawn that belong to the hyperplane in question.
In an alternative view furnished by the remaining three pictures, instead of being stacked on top of each other, the doily-quads are put side by side to make 
finer traits of the structure more discernible.
The points and lines of a hyperplane are boldfaced. If the point is encircled, then the type-one line passing through it is fully
contained in the hyperplane; the doubled (type-two) lines are those which belong to a deep grid-quad. Note that singular hyperplanes are those of type $H_2$.}
\end{figure}

\begin{figure}[pth!]
\centerline{\includegraphics[width=12.cm,clip=]{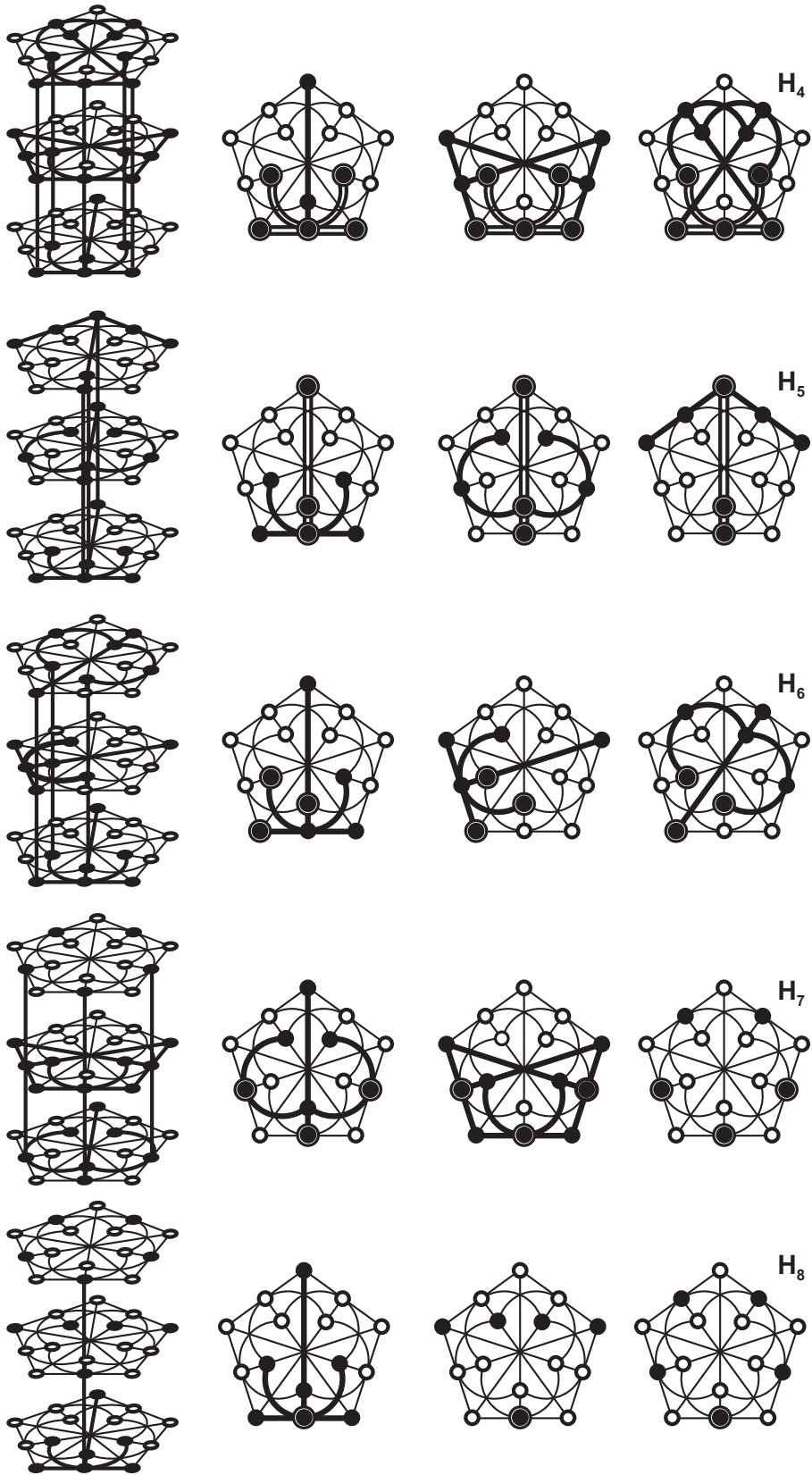}}
\vspace*{.2cm} \caption{A diagrammatic illustration of the structure of the five kinds of geometric hyperplanes of the second family with the same symbols/notation as in the preceding figure.}
\end{figure}

The cardinalities of hyperplanes thus sum up to $93 + 930 = 1023 = 2^{10} - 1 = |$PG(9,\,2)$|$.
This also proves that we have enumerated all hyperplanes since any slim dense near hexagon is universally embeddable into a projective space over GF(2) and the geometric hyperplanes are 
precisely those arising from its universal embedding \cite{ron}, and that in our particular case --- as already mentioned in Section 3.1 --- the universal
embedding is indeed into PG(9,\,2) \cite{bchw}.

Table 2 reveals a number of interesting facts. First, one readily observes that $|H_i|$ = 1 (mod 4) for any $i$ as per their {\it point} cardinality.\footnote{In this respect our near hexagon resembles the {\it dual} of the split Cayley hexagon of order two, where $|H_i|$ = 3 (mod 4), any $i$ \cite{fj}.} 
Next, there is no hyperplane featuring points of every order. Similarly, there is no hyperplane endowed with all
the four kinds of doily-quads. On the other hand, there are two distinct types of hyperplane ($H_4$ and $H_5$) containing all the three kinds of grid-quads and every single hyperplane contains a singular grid-quad. Interestingly, there are a couple
of types of hyperplane devoid of deep points ($H_6$ and $H_7$) and the same number of those having isolated points (that is, points of order zero --- $H_7$ and $H_8$). Furthermore, there are  as many as 
three distinct line types
having identical point/line cardinality ($H_5$ to $H_7$). It is also worth mentioning that the complement of an $H_1$ is a pair of dual grids, GQ(1,\,2)s --- a distant relative of Schl\" afli's double-six.

We conclude this section with the following observation. Picking up a grid in one of the three doilies, there is a unique associated slim dense near hexagon of type L$_3$\,$\times$\,GQ(2,\,1) $\simeq$ 
L$_3^{\times 3}$ sitting inside L$_3$\,$\times$\,GQ(2,\,2). This (smallest slim dense) near hexagon features five distinct kinds of geometric hyperplanes \cite{gs} and viewing it as embedded in L$_3$\,$\times$\,GQ(2,\,2), 
it can easily be demonstrated that each of them arises from a hyperplane of L$_3$\,$\times$\,GQ(2,\,2); in fact, one can prove a stricter condition, namely that every hyperplane of L$_3$\,$\times$\,GQ(2,\,1)
originates from one of the hyperplanes of type $H_2$, $H_5$ or $H_6$, because the totality of these span a subspace of PG(9,\,2) isomorphic to PG(7,\,2), and so to $\mathcal{V}(L_{3}^{\times 3})$ as well (see \cite{gs}). We also note in passing that since two distinct copies of a grid in a doily always meet in a pair of concurrent lines
(see Figure 2, top row),
two different L$_3$\,$\times$\,GQ(2,\,1)'s of L$_3$\,$\times$\,GQ(2,\,2) share a pair of concurrent grid-quads.

\section{Conclusion}
We have given a detailed description of all different types of geometric hyperplanes of the point-line incidence geometry L$_3$\,$\times$\,GQ(2,\,2), which is the smallest slim dense near hexagon
featuring two distinct kinds of quads --- namely grid-quads and doily-quads. The hyperplanes, whose total number amounts to $1023 = 2^{10} - 1 = |$PG(9,\,2)$|$, are of eight types and they form two distinct families according as
they contain a deep doily-quad or not. This two-family split was demonstrated to have a natural explanation in terms of the points and lines of the Veldkamp space of the
generalized quadrangle of order two. Each hyperplane's type is uniquely characterized by the following string of parameters (Table 2): the number of points  and lines of a representative, followed by the cardinalities of the points
of a given order, cardinalities of deep, singular, ovoidal and subquadrangular quads of both kinds, and, finally, by the total number of its copies. Several interesting combinatorial properties were
also explicitly mentioned. 

We believe that L$_3$\,$\times$\,GQ(2,\,2), like GQ(2,\,2) itself, will play a prominent role in the context of both quantum information theory and entropy formulas of some yet unknown stringy black hole/ring
solutions. It is especially the former domain when we surmise that the combinatorics and geometry of L$_3$\,$\times$\,GQ(2,\,2) could mimic a whole class of {\it three two-qubit} systems entangled in a particular way and lead to the notion of a ``twisted'' Mermin square. In this respect, a particularly attractive task to address is as follows. It is a well established fact \cite{spp,ps2} that the structure of GQ(2,\,2) underlies the commutation relations
between the 15 operators of two-qubit Pauli group, and that grids sitting in it generate Mermin magic squares. Now, suppose that we label all the three doilies of L$_3$\,$\times$\,GQ(2,\,2) by
two-qubit Pauli matrices. We start from the configuration in which the labels on each line of type one are the same. Then we keep labels of one of the doilies fixed and let the group $S_6$ act on the labels/points of the other two. 
An interesting question emerges: how many Mermin squares can we get among {\it grid-quads} and how are they coupled to each other? (In our starting position there are none.)
L$_3$\,$\times$\,GQ(2,\,2) may even turn out to be of relevance for three-qubits as its automorphism group, $S_6 \times S_3$, is isomorphic to a maximal subgroup of
$W'(E_7)$ that can be given an ``entangled" three-qubit representation \cite{pla}.  Explorations along these lines are already well under way and will be dealt with in a separate paper.

From a mathematical point of view, as the very next step it is desirable to find the stabilizer group for a representative of each hyperplane's type and the corresponding point orbit sizes. Then, we shall embark on examining all the types of Veldkamp lines of $\mathcal{V}$(L$_3$\,$\times$\,GQ(2,\,2)); since, obviously, $\mathcal{V}$(L$_3$\,$\times$\,GQ(2,\,2)) $\simeq$ PG(9,\,2), this will
be a much more demanding task because PG(9,\,2) is endowed with as many as 174\,251 lines. 

\vspace*{.5cm} 
\noindent
{\bf Acknowledgements}\\
\normalsize This work was carried out within the framework of the Cooperation Group
``Finite Projective Ring Geometries: An Intriguing Emerging Link Between Quantum Information Theory, Black-Hole Physics and Chemistry of Coupling''  at the Center for Interdisciplinary Research (ZiF),
University of Bielefeld, Germany.  M. S. was also partially supported by the VEGA grant
agency projects Nos. 2/0092/09 and 2/7012/27. We are extremely grateful to Harm Pralle (TU Clausthal) for making our classification of hyperplanes rigorous and to Richard Green (University of Colorado, Boulder) for a number of enlightening comments.

\end{document}